\documentstyle[12pt]{article}
\begin{document}
\parskip 10pt plus 1pt
\title{Vacuum Expectation Value of the Higgs field and Dyon Charge Quantisation
from Spacetime Dependent Lagrangians}
\author{
{\it Rajsekhar Bhattacharyya}$^{*}$\\
{Dept.of Physics, Dinabandhu Andrews College, Kolkata-700084, INDIA}\\
{\it  Debashis Gangopadhyay$^{\dagger}$}\\ 
{S.N.Bose National Centre For Basic Sciences}\\
{JD Block, Sector-III, Salt Lake, Kolkata-700098, INDIA}\\
}
\date{}
\maketitle
\baselineskip=20pt
\begin{abstract}
The spacetime dependent lagrangian formalism of references
[1-2] is used to obtain a classical solution of Yang-Mills theory. This
is then used to obtain an estimate of the vacuum expectation value of the Higgs
field , {\it viz.} $\phi_{a}= A/e$, where $A$ is a constant and $e$ is the 
Yang-Mills coupling (related to the usual electric charge). The solution can
also  accommodate non-commuting coordinates on the boundary of the theory which
may be used to construct $D-$brane actions.The formalism is also used to 
obtain the Deser-Gomberoff-Henneaux-Teitelboim results [10] for
dyon charge quantisation in abelian $p$-form theories in dimensions $D=2(p+1)$ 
for both even and odd p.

\end{abstract}
\newpage
The spacetimetime dependent lagrangian formalism [1-2] gives an alternative
way to deal with electromagnetic duality [3], weak-strong duality [4] and electro-
gravity duality [5]. Here this method will be used (a)to obtain an estimate of the
vacuum expectation value (VEV) of the Higgs field in terms of the electric charge
$e$ and a constant (b)to show that the 't Hooft ansatz for
obtaining the "t Hooft-Polyakov monopole solution is sufficiently general to
lead to other solutions containing coordinates near the boundary that do 
not commute (c)to show that the 't Hooft ansatz for the gauge 
field is sufficient to yield a solution for the Higgs field for 
$r\rightarrow\infty$ without the necessity of any further ansatz for $\phi$ and
(d)to obtain the results of Deser,Gomberoff,Henneaux and Teitelboim [10] 
regarding dyon charge quantisation in abelian , $p-$form theories.Results (a) to(c) will be obtained in Section 1 while Section 2 Deals with (d). 

{\bf 1. VEV of Higgs field }\\
In [1] it was shown that if 
the lagrangian $L'$  be a function of fields $\eta_{\rho}$, 
their derivatives $\eta_{\rho,\nu}$ {\it and the spacetime 
coordinates $x_{\nu}$}, and $L'$ be written as
$L'(\eta_{\sigma},\eta_{\sigma,\nu},.. x_{\nu})
=L(\eta_{\sigma},\eta_{\sigma,\nu})\Lambda(x_{\nu})$, then 
the variational principle [12] gives equations of motion as
$$\int dV \left(\partial_{\eta}(L\Lambda)
- \partial_{\mu}\partial_{\partial_{\mu}\eta}(L\Lambda)\right) = 0\eqno(1)$$
$\Lambda(x_{\nu})$ is the $x_{\nu}$ dependent part in $L'$ and is a finite 
non-dynamical and non-vanishing function.
It is like an external field and equations of motion for $\Lambda$ meaningless. 
Duality invariance is related to finiteness of $\Lambda$ on the boundary. When equations of
motion are duality invariant, finiteness of $\Lambda$ on the spatial boundary
at infinity leads to new solutions for the fields.  The finite behaviour of $\Lambda$ on the boundary encodes the exotic solutions of the theory within the boundary thus  reminding one of the holographic principle[9]. 

Consider the Georgi-Glashow model with [3]
$$L= [-(1/4)G^{\mu\nu}_{a} G_{a\enskip\mu\nu} + (1/2)(D^{\mu}\phi)_{a}(D_{\mu}\phi)_{a} - V(\phi)]\eqno(2a)$$
where usually one takes $V(\phi)= (\lambda/4)(\phi^{a}\phi^{a}-a^{2})^{2}$.
The gauge group is $SO(3)$,$a,b,c$ are $SO(3)$ indices, with the generators
$\tau^{a}$ satisfying $[\tau^{a},\tau^{b}]=i\epsilon^{abc}\tau^{c}$.Gauge fields
$W_{\mu}=W_{\mu}^{a}\tau^{a}$. The field strength is
$G^{\mu\nu}_{a}= \partial^{\mu}W^{\nu}_{a} 
- \partial^{\nu}W^{\mu}_{a} - e\epsilon_{abc}W^{\mu}_{b}W^{\nu}_{c}$;
$\tilde G^{\mu\nu}_{a}=(1/2)\epsilon^{\mu\nu\rho\sigma}G_{a\enskip\rho\sigma}$;
and the matter fields $\phi$ are in the adjoint representation of $SO(3)$.
The energy density $\Theta_{00}=(1/2)[(E^{i}_{a})^{2}+(B^{i}_{a})^{2}+(D^{0}\phi_{a})^{2}
+(D^{i}\phi_{a})^{2}+ V(\phi)]\geq 0$.
The non abelian electric and magnetic fields are defined respectively as:
$E^{i}_{a}=-G^{0i}_{a}$  and  $B^{i}_{a}=-(1/2)\epsilon^{i}_{jk}G_{a}^{jk}$.
The vacuum configuration is
$G^{\mu\nu}_{a}=0\enskip;\enskip D_{\mu}\phi=0\enskip;\enskip V(\phi)=0$.
There are also the  Bianchi identities $D^{\mu}\tilde G_{a\enskip\mu\nu}=0$.
Duality invariance means that $D^{\mu}G_{a\enskip\mu\nu}=0$.

In [1] it was shown that for this theory  
$$L'= L\Lambda=[-(1/4)G^{\mu\nu}_{a} G_{a\enskip\mu\nu} 
+ (1/2)(D^{\mu}\phi)_{a}(D_{\mu}\phi)_{a} 
- V(\phi)]\Lambda(x_{\nu})\eqno(2b)$$
The equations of motion using $(1)$ were :
$$\Lambda (D^{\mu}G_{a\enskip\mu\nu}) + (\partial^{\mu}\Lambda)G_{a\enskip\mu\nu}
+ \Lambda e \epsilon_{abc} (\partial_{\nu}\phi)_{b}(\phi)_{c} 
- \Lambda e^{2} \epsilon_{abc}\epsilon_{bc'd'}W_{\nu\enskip c'} 
\phi_{c} \phi_{d'} = 0\eqno(3a)$$   
$$(D^{\mu}D_{\mu}\phi)_{a}\Lambda + (D_{\mu}\phi)_{a}\partial_{\mu}\Lambda 
= - (\partial_{{\phi}^{a}}V)\Lambda\eqno(3b)$$
and the Bianchi identities were: $D^{\mu}\tilde G_{a\enskip\mu\nu}=0$
Requiring duality invariance (i.e.$D^{\mu}G_{a\enskip\mu\nu}=0$) gave the solution to $\Lambda\equiv\Lambda(r)$ as
$$\Lambda_{\infty} 
= \Lambda_{0} exp[-e \int_{0}^{\infty} dr \left( (\epsilon_{abc} 
(D_{\nu}\phi)_{b} \phi_{c}) 
(\partial^{i} r G_{a \enskip i\nu})^{-1}\right)]\eqno(4)$$
where $\Lambda_{p}$ is the value of $\Lambda$ at $r=p$;
$a,\nu$ are fixed; and there is a sum over indices $i,b$ and $c$.
$\Lambda_{\infty}$ must be finite. Choose this
to be the constant unity.This may be realised in various ways, the simplest being
$(D_{\nu} (\phi)_{b} \Rightarrow 0$, \enskip 
$(\phi)_{c} \Rightarrow finite$, \enskip
and the product
$(D_{\nu}\phi)_{b} (\phi)_{c}$ falls off faster than $G_{a \enskip i\nu}$ for
large $r$. Then a constant value for $\Lambda$ was perfectly consistent with $(3b)$ and
the conditions became analogous to the Higgs' vacuum condition for
the t'Hooft-Polyakov monopole solutions where the duality invariance of the 
equations of motion and Bianchi identities are attained at large $r$ by 
demanding $(D_{\mu} \phi)_{a} \Rightarrow 0$ and $\phi_{a}\Rightarrow a\delta_{a3}$
at large $r$. The results were perfectly consistent with the 
usual choice for the Higgs' potential $V(\phi)$ even though nothing had been 
assumed regarding this. So the t'Hooft-Polyakov monopole solutions 
followed naturally in our formalism. We now discuss two other interesting possibilities.

{\bf Case I}

$$\left(\epsilon_{abc} (D_{\nu}\phi)_{b}\phi_{c}\right) \Rightarrow 0\eqno(5)$$ 
(i.e. the duality condition $(D^{\mu}G_{\nu\mu})_{a}=0$)
and falls off faster than $G_{a \enskip i\nu}$ for large $r$ 
($a$ and $\nu$ are fixed). A solution is when 
$$D_{\nu}  \phi = \alpha_{\nu} \phi\eqno(6)$$
where $\alpha_{\nu}$ can be any Lorentz four vector field that is consistent with all 
the relevant equations of motion and the minimum
energy requirements. The minimum energy requirements are satisfied because it
is straightforward to verify that the gauge fields $W^{\mu}_{a}$ do not change.
This is seen by taking
the cross product of $\phi$ ($\phi$ is a $SO(3)$ vector)
with equation $(6)$. We again arrive at the well known results of Corrigan 
{\it et al} [7], {\it viz.} $W^{\mu}= (1/a^{2}e)\phi\wedge\partial^{\mu}\phi +(1/a)\phi A^{\mu}$,
where $A^{\mu}$ is arbitrary.

As the gauge fields $W^{\mu}_{a}$ do not change, so even with this solution 
we obtain the same gauge field solutions as before and so minimum energy 
requiremet is automatically satisfied. However, this new solution allows us 
to obtain an estimate of the vacuum expectation value of the Higgs field and
to this we now proceed. Let 
$\alpha_{\nu}=(0,\alpha_{i})\equiv \alpha(r) \hat r$, where $\hat r$ is the unit
radial vector.So the Bogomolny condition is $B^{i}_{a}=D^{i}\phi_{a}=\alpha^{i}\phi_{a}$ 
and the Higgs vacuum condition obtained from equation $(3b)$ 
(for $r\rightarrow \infty, \Lambda$ is a constant , say unity) is 
$$[D^{i}(\alpha_{i}\phi)]_{a}=-\partial_{\phi^{a}} V = 0\eqno(7a)$$
i.e. we are at a minima of the potential $V$. If $\phi_{a}\not= 0$, $(7a)$ implies
$$div\enskip \vec{\alpha} + \vec{\alpha}^{2} =0\eqno(7b)$$
and the solution is 
$$\alpha(r)= 1/(cr^{2}-r)\enskip , \alpha^{i}=r^{i}/(cr^{3}-r^{2})\eqno(8)$$
where we take the constant $c$ to be negative.
Let us now take the 't Hooft ansatz for the gauge field, {\it viz.}
$$W^{0}_{a}=0\enskip \enskip ;  W^{i}_{a}= -\epsilon_{aik}r^{k}[1-K(aer)]/(er^{2})\eqno(9)$$
where the function $K(aer)$ has been well studied [3] and goes to zero at $r\rightarrow\infty$.
Then  the electric field vanishes while $G_{a\enskip jk}, B^{i}_{a}$ are
$$G_{a\enskip jk}
=(1/er^{2})[2\epsilon_{ajk}(1-K)
+\epsilon_{akl}r^{l}\partial_{j}K-\epsilon_{ajl}r^{l}\partial_{k}K]$$
$$+(1/er^{4})[2(1-K)(\epsilon_{akl}r^{l}r_{j}-\epsilon_{ajl}r^{l}r_{k})
+(1-K)^{2}(\delta_{aj}\epsilon_{ckl}r^{c}r^{l}-\epsilon_{jkl}r_{a}r^{l})]\eqno(10a)$$
$$B^{i}_{a}= (1/er^{4})[(1-K)^{2}r_{a}r^{i}-2(1-K)r^{i}r_{a}]
-(1/er^{2})[r^{i}\partial_{a}K-\delta^{i}_{a}r^{m}\partial_{m}K]\eqno(10b)$$
Now $B^{i}_{a}=D^{i}\phi_{a}=\alpha^{i}\phi_{a}$.Therefore,taking $c=-A$ so that $A$
is positive we have
$$\phi_{a}={(1+Ar)\over{(er^{2})}}[2(1-K)r_{a}-(1-K)^{2}r_{a}+r^{2}\partial_{a}K-r_{a}r^{m}\partial_{m}K]\eqno(10c)$$
It is easily seen that $(10c)$ reduces to the 't Hooft ansatz for $\phi_{a}$
for $A=0$ and $r\rightarrow\infty$. Thus we have obtained an expression for
$\phi$ without assuming any ansatz. This had never been possible before.
There is another interesting outcome.For $r\rightarrow\infty$ we have
$K\rightarrow 0$ and so 
$$\phi_{a}\rightarrow {Ar^{a}\over er} + {r^{a}\over er^{2}}
={A\over {e}}\hat r^{a} +{\hat r\over {er}}
\rightarrow{A\over {e}} \hat r^{a}\eqno(11)$$
for $r\rightarrow\infty$. But 
$\phi_{a}=a\delta_{a3}$ for $r\rightarrow\infty$. Therefore $a=A/e$. Thus we have obtained the VEV of the 
Higgs field in terms of $e$.

{\bf Case II}

$\alpha_{\nu}$ is any Lorentz four vector field as in {\bf I} but which 
may also carry internal symmetry indices {\it other than $SO(3)$}
with the generators of the symmetry satisfying some Lie algebra 
$[T_{P},T_{Q}]= if_{PQR} T_{R}$. Let us take the group to be $SU(2)$.
i.e. say, $\alpha^{\nu}= \alpha^{\nu}_{P}T_{P}$; $P,Q,R=1,2,3$;
$T_{P}$ being the generators of $SU(2)$. 
Again choosing $\alpha^{\nu}=(0,\alpha^{i}_{P}T^{P})$ with
$\vec\alpha_{P}=\alpha_{P}(r)\hat r$ and
using the well known properties of 
the Pauli matrices it is easily seen that  the analogue of equation $(7b)$ is
$$div\enskip \vec{\alpha_{P}}=0\eqno(12)$$
which has the solution
$$\alpha^{i}_{P}=A_{P} {r^{i}\over {r^{3}}}\eqno(13)$$
where $A_{P}$ are constants. Writing $r^{i}_{P}=A_{P} r^{i}=r^{3}\alpha^{i}_{P}$,
we can then define new coordinates 
$$R^{i}=r^{i}_{P}T_{P}\enskip ; [R^{i},R^{j}]\not=0\eqno(14)$$
and these are non-commuting. Moreover, they carry both Lorentz and internal
indices and hence are like gauge fields in some different theory. Note that 
transverse coordinates (i.e. transverse to the brane and lying in the bulk volume)
in $D$ brane theories are often identified with gauge
fields [8] and so we can construct such actions with our solutions $(14)$.
Under these circumstances, equation  $(6)$ should be written as 
$$\partial_{\mu}^{P}\phi_{a}-e\epsilon_{abc}(W_{\mu}^{b})^{P}\phi_{c}=\alpha_{\mu}^{P}\phi_{a}$$
where the coordinates and their differentials are now matrices and capital
alphabets denote the indices of the new symmetry group. For fixed $P$,  
$(W_{\mu}^{b})^{P}$ may be identified with the old gauge fields $W_{\mu}^{b}$.

A point to note is that we have taken the symmetry group for $\alpha_{\nu}$ to be 
some group other than $SO(3)$. This is to ensure in the simplest possible way 
that the fields $(W_{\mu}^{b})^{P}$ {\it for fixed P} may be identified with the
old (i.e.unchanged) gauge fields $W_{\mu}^{b}$ ($P$ is now a fixed index) and 
so the minimum energy requirements are satisfied in each sector of $P=1,2,3$.
(This is seen by taking the cross product of $\phi$ with the analogue of 
equation $(6)$ and proceeding as before).So each sector now contains a monopole.
Then we have a configuration that is quite similar to "string" of monopole 
solutions connecting two D-branes. Such configurations are known in the literature [8].
The other point is that the vacuum expectation value of the Higgs field is 
proportional to the inverse of the coupling $e$; and this result has been 
obtained from the classical solutions. This result is similar to that obtained
in Ref.[2b] if we are ready to identify the inverse of Newton's gravitational
constant (which is definitely the coupling constant in theories of gravity)
as the vacuum expectation value of some field hitherto unknown.

The solutions in equation $(6)$ were hidden in 't Hooft-Polyakov's 
work. This had been overlooked before for the simple reason because at that point
of time one was more concerned in obtaining solutions from the minimum 
(finite ) energy principles. This was perfectly justified. We have obtained
the solutions  from the requirement of duality invariance which is quite relevant
at this point of time. However, we have also shown that the duality requirements
automatically contain the minimum energy condition ($\Lambda$ is also finite for $D_{\nu}\phi=0$).
All the results have been obtained at $r\rightarrow\infty$. That is, we are at the boundary 
of the theory. So the finiteness of $\Lambda$ at the boundary encodes the duality
invariance of the theory within the boundary and thus an analogue of the 
holographic principle [9] seems to be at work. On the boundary there seems to
exist {\it a different gauge field theory} together with non-commuting coordinates.

{\bf 2. Dyon charge quantisation in abelian $p$-form theories}

In [1] the formalism of spacetime dependent lagrangians was used to obtain 
the Dirac quantisation condition. Here we shall follow the same method to obtain
the results of Deser,Gomberoff,Henneaux and Teitelboim [10] regarding dyon
charge quantisation in abelian , $p-$form theories. Our results will be 
obtained from a simple generalisation of the lagrangian constructed in [1] and
a generalisation of the interaction terms using the completely
antisymmetric symmetric tensor 
$\epsilon_{ab}=\left(\matrix{0&1\cr -1&0\cr}\right)$
and the symmetric tensor
$\rho_{ab}=\left(\matrix{0&1\cr 1&0\cr}\right)$.

In[1] we considered a $U(1)\otimes U(1)$ gauge invariant theory.$A_{\mu}$ and 
$B_{\mu}$ were four-vector potentials corresponding to electric ($e$)
and magnetic ($g)$ charges; $F_{\mu\nu}$, $G_{\mu\nu}$ were the respective 
field strengths; $j_{\mu}$, $k_{\mu}$ were the electric and magnetic (current)
sources with interactions between respective currents and potentials introduced
in the usual way
$$L_{1}=-(1/4)F^{\mu\nu}F_{\mu\nu}-(1/4)G^{\mu\nu}G_{\mu\nu}-j^{\mu}A_{\mu}-k^{\mu}B_{\mu}\eqno(15a)$$
with $F^{\mu\nu}=\partial^{\mu}A^{\nu}-\partial^{\nu}A^{\mu}\enskip ; 
G^{\mu\nu}=\partial^{\mu}B^{\nu}-\partial^{\nu}B^{\mu}\enskip ;
\tilde G^{\mu\nu}=(1/2)\epsilon^{\mu\nu\rho\sigma}G_{\rho\sigma}\enskip ; 
\partial^{\mu}j_{\mu}=\partial^{\mu}k_{\mu}=0$ (current conservation)$\enskip;
\partial^{\mu}A_{\mu}=\partial^{\mu}B_{\mu}=0$ (transversality)$\enskip ;
\partial^{\mu}F_{\mu\nu}=j_{\nu}\enskip ; \partial^{\mu}\tilde F_{\mu\nu}=0\enskip ;
\partial^{\mu}G_{\mu\nu}=k_{\nu}\enskip ; \partial^{\mu}\tilde G_{\mu\nu}=0$.
Defining (note that $\tilde{\tilde F}=-F$ and  $\tilde{\tilde G}=-G$)
$$\xi^{\mu\nu}=F^{\mu\nu}+\tilde G^{\mu\nu}\enskip ;
\tilde\xi^{\mu\nu}=F^{\mu\nu}-\tilde G^{\mu\nu}\eqno(15b)$$
means $\partial^{\mu}\xi_{\mu\nu}=j_{\nu}\enskip ; \partial^{\mu}\tilde\xi_{\mu\nu}=-k_{\nu}$.
A complex interaction term $if(\Lambda)\alpha A^{\mu}B_{\mu}j^{\nu}k_{\nu}$
was introduced where $f(\Lambda)$ was a dimensionless function of $\Lambda$,
and the spacetime dependent lagrangian was written as
$$L=[-(1/4)F^{\mu\nu}F_{\mu\nu}-(1/4)G^{\mu\nu}G_{\mu\nu}-j^{\mu}A_{\mu}-k^{\mu}A_{\mu}
+if(\Lambda)\alpha A^{\mu}B_{\mu}j^{\nu}k_{\nu}]\Lambda(x)\eqno(15c)$$
Equations of motion using $(1)$ were set up, duality invariance imposed
and the solution for $\Lambda$ obtained for appropriate sources $j_{\mu},k_{\mu}$.
Finiteness of $\Lambda$ at $r\rightarrow\infty$
led to the Dirac quantisation condition. The $U(1)\otimes U(1)$ invariance
of the original theory was broken.

We now use the above procedure to obtain the dyon charge quantisation condition 
for abelian $p$-form theories. First consider dimension $D=4$. Then $p=1$.
There are now two objects, each of which carries both electric ($e$) and magnetic
($g$) charges.Accordingly, there will be two $F$ 's, two $G$ 's
two $A$ 's, two $B$ 's and two $j$ 's and two $k$ 's.
Let the index $a=1,2$ denote this. We next choose the interaction term as 
$if(\Lambda)\epsilon^{bc}\alpha A^{\mu}_{a}B_{a\enskip\mu} j^{\nu}_{b} k_{c\enskip\nu}$.
Then the generalisation of the lagrangian $(15a)$ becomes
$$L = [-(1/4) F^{\mu\nu}_{a} F_{a\enskip\mu\nu} - (1/4) G^{\mu\nu}_{a} G_{a\enskip\mu\nu}
- j^{\mu}_{a} A_{a\enskip\mu} - k^{\mu}_{a} B_{a\enskip\mu}$$
$$+ i\enskip f(\Lambda)\enskip \alpha\enskip \epsilon^{bc}\enskip 
A^{\mu}_{a}\enskip B_{a\enskip\mu}\enskip j^{\nu}_{b}\enskip k_{c\enskip\nu}]\enskip \Lambda(x)\eqno(16)$$
As before $\xi^{\mu\nu}_{a}=F^{\mu\nu}_{a}+\tilde G^{\mu\nu}_{a}\enskip ;
\tilde\xi^{\mu\nu}_{a}=F^{\mu\nu}_{a}-\tilde G^{\mu\nu}_{a}$. Equations of 
motion that follow from $(1)$ are (for each $a=1,2$):
$$\Lambda(\partial^{\mu}\xi_{a\enskip\mu\nu})+[(\partial^{\mu}\Lambda)F_{a\enskip\mu\nu}
-\Lambda(j_{a\enskip\nu}+ic_{a\enskip\nu})]=0\eqno(17a)$$
$$\Lambda(\partial^{\mu}\tilde\xi_{a\enskip\mu\nu})-[(\partial^{\mu}\Lambda)G_{a\enskip\mu\nu}
-\Lambda(k_{a\enskip\nu}+id_{a\enskip\nu})]=0\eqno(17b)$$
where $c_{a\enskip\nu}=f(\Lambda) \alpha \epsilon^{bc} j^{\mu}_{b} k_{c\mu} B_{a\enskip\nu}\enskip;
d_{a\enskip\nu}=f(\Lambda) \alpha \epsilon^{bc} j^{\mu}_{b} k_{c\mu} A_{a\enskip\nu}$.
Duality invariance means 
$\partial^{\mu}\xi_{a\enskip\mu\nu}=0$\enskip and 
$\partial^{\mu}\tilde\xi_{a\enskip\mu\nu}=0$. This therefore implies
$$(\partial^{\mu}\Lambda)F_{a\enskip\mu\nu}
-\Lambda(j_{a\enskip\nu}+ic_{a\enskip\nu})=0\eqno(18a)$$
$$(\partial^{\mu}\Lambda)G_{a\enskip\mu\nu}
-\Lambda(k_{a\enskip\nu}+id_{a\enskip\nu})=0\eqno(18b)$$
To solve the above for specific sources we take
$j^{\nu}_{a}=e_{a}\int dx^{\nu}\delta^{4}(x)\enskip;
k^{\nu}_{a}=g_{a}\int dx^{\nu}\delta (x_{3}-b)\delta^{3}(x)$. Now assume 
$\Lambda=\Lambda(x_{3})$ and
that only the $\nu=0$ component of the sources are present so that
$j^{0}_{1}=e_{1}\delta (x_{1}) \delta (x_{2}) \delta (x_{3}) \enskip;
k^{0}_{1}=g_{a}\delta (x_{1})\delta (x_{2}) \delta (x_{3})$ and 
$j^{0}_{2}=e_{2}\delta (x_{1}) \delta (x_{2}) \delta (x_{3-b}) \enskip;
k^{0}_{2}=g_{2}\delta (x_{1})\delta (x_{2}) \delta (x_{3-b})$. 
Then we get for $\nu=0,1,2$
$$(\partial^{3}\Lambda)F_{a\enskip 3\nu}=\Lambda(j_{a\enskip\nu}+ic_{a\enskip\nu})\eqno(18a)$$
$$(\partial^{3}\Lambda)G_{a\enskip 3\nu}=\Lambda(k_{a\enskip\nu}+id_{a\enskip\nu})\eqno(18b)$$
For $\nu=3,F_{a\enskip 33}=G_{a\enskip 33}=0$ for all $a$, and the solutions to $(18a)$ and $(18b)$ for $\nu=0$ are:
$$\Lambda_{\infty}\enskip=\Lambda_{-\infty}exp[e_{a}\delta(x_{1})\delta(x_{2})/F_{a\enskip 30}(x_{0},x_{1},x_{2},0)]$$
$$exp[if(\Lambda)\alpha\epsilon_{bc}e_{b}g_{c}P_{a\enskip 0}(x_{0},x_{1},x_{2},b)]$$
$$=\Lambda_{-\infty}exp[e_{a}\delta(x_{1})\delta(x_{2})/F_{a\enskip 30}(x_{0},x_{1},x_{2},0)]$$
$$exp[if(\Lambda)\alpha (e_{1}g_{2}-e_{2}g_{1})P_{a\enskip 0}(x_{0},x_{1},x_{2},b)]\eqno(19a)$$
$$\Lambda_{\infty}\enskip=\Lambda_{-\infty}exp[g_{a}\delta(x_{1})\delta(x_{2})/G_{a\enskip 30}(x_{0},x_{1},x_{2},0)]$$
$$exp[if(\Lambda)\alpha\epsilon_{bc}e_{b}g_{c}Q_{a\enskip 0}(x_{0},x_{1},x_{2},b)]$$
$$=\Lambda_{-\infty}exp[e_{a}\delta(x_{1})\delta(x_{2})/G_{a\enskip 30}(x_{0},x_{1},x_{2},0)]$$
$$exp[if(\Lambda)\alpha (e_{1}g_{2}-e_{2}g_{1})Q_{a\enskip 0}(x_{0},x_{1},x_{2},b)]\eqno(19b)$$
$$P_{a\enskip 0}(x_{0},x_{1},x_{2},b)=(\delta (x_{1}))^{2})(\delta (x_{2}))^{2}\delta (b)B_{a0}(x_{0},x_{1},x_{2},b)/F_{a\enskip 30}(x_{0},x_{1},x_{2},b)\eqno(20a)$$
$$Q_{a\enskip 0}(x_{0},x_{1},x_{2},b)=(\delta (x_{1}))^{2})(\delta (x_{2}))^{2}\delta (b)A_{a0}(x_{0},x_{1},x_{2},b)/G_{a\enskip 30}(x_{0},x_{1},x_{2},b)\eqno(20b)$$
Proceeding as in ref.[1],
choose $\Lambda_{\infty}=\Lambda_{-\infty}=1$ and consider the set of equations
$(8a)$ and $(9a)$. The two exponentials must reduce to unity. For the first
exponential this implies the Dirac string configuration where $F_{a\enskip 30}\rightarrow\infty$,
and so the exponential becomes unity. For the second exponential, the numerator
in $(9a)$ has singular $\delta$-functions and together with $B_{a\enskip 30}\rightarrow\infty$
since $F_{a\enskip 30}\rightarrow\infty$.So second exponential is unity if 
$exp[if(\Lambda)\alpha(e_{1}g_{2}-e_{2}g_{1})P_{a\enskip 0}]=1$,i.e.
$exp[if(\Lambda)\alpha(e_{1}g_{2}-e_{2}g_{1})]^{P_{a\enskip 0}}=1$ (as $P_{a\enskip 0}$
is finite). Therefore
$$f(\Lambda)\alpha(e_{1}g_{2}-e_{2}g_{1})=2\pi n\eqno(10)$$
All the above results are true in each sector, {\it viz.}, $a=1,2$.
As in ref. [1], there are two possibilities: (a)$f(\Lambda)=0$. Then the 
$U(1)\otimes U(1)$ invariance in each sector of $L$ is unbroken and we have
the Dirac string configuration from the first exponential $F_{a\enskip 30}\rightarrow\infty$.
(b)$f(\Lambda)= a\enskip finite\enskip constant$. Then the 
$U(1)\otimes U(1)$ invariance in each sector of $L$ is broken and putting
$\alpha=(\hbar)^{-1}$, {\it we get the Dirac quantisation condition for dyons.}
For $\nu=1,2$ a similar analysis will again lead to $(10)$ 
and similarly for the set of equations $8(b)$ and $(9b)$. Note that we have taken
the same function $\Lambda$ in each sector of the theory. This is justifiable
from the fact that finally we proceed to the case of $\Lambda$ becoming unity
in each sector.

Now consider dimension $D=6$. This means $p=2$. So we have 2-form potentials $A_{a}^{\mu\nu}\enskip ;\enskip B_{a}^{\mu\nu}$.
Then each of the antisymmetric field strengths $F , G$ will be a 3-form in the Lorentz indices
and we have the following constructions:
$$3-form\enskip field\enskip stengths:\enskip F_{a}^{\mu\nu\sigma}=\partial^{\mu}A_{a}^{\nu\sigma}- \partial^{\nu}A_{a}^{\mu\sigma}+ \partial^{\sigma}A_{a}^{\mu\nu}$$
$$\enskip\enskip\enskip\enskip\enskip\enskip\enskip\enskip\enskip\enskip G_{a}^{\mu\nu\sigma}=\partial^{\mu}B_{a}^{\nu\sigma}- \partial^{\nu}B_{a}^{\mu\sigma}+ \partial^{\sigma}B_{a}^{\mu\nu}$$
$$\xi^{\mu\nu\sigma}_{a}=F^{\mu\nu\sigma}_{a}+\tilde G^{\mu\nu\sigma}_{a}\enskip ;
\enskip\tilde\xi^{\mu\nu\sigma}_{a}=F^{\mu\nu\sigma}_{a}-\tilde G^{\mu\nu\sigma}_{a}$$
$$2-form\enskip antisymmetric\enskip potentials\enskip A_{a}^{\mu\nu}\enskip;\enskip B_{a}^{\mu\nu}\enskip (antisymmmetric\enskip w.r.t.\enskip \mu,\nu) $$
$$Dual:\enskip\tilde F_{a}^{\mu\nu\sigma}=(1/3!)\epsilon^{\mu\nu\sigma\alpha\beta\gamma} F_{a\enskip\alpha\beta\gamma}\enskip ;\enskip\tilde G_{a}^{\mu\nu\sigma}=(1/3!)\epsilon^{\mu\nu\sigma\alpha\beta\gamma} G_{a\enskip\alpha\beta\gamma}$$
$$2-form\enskip currents:\enskip j_{a}^{\mu\nu}=e_{a}\int dx^{\mu}\wedge dx^{\nu}\delta^{6}(x)\enskip;
\enskip k_{a}^{\mu\nu}=g_{a}\int dx^{\mu}\wedge dx^{\nu}\delta^{6}(x)$$
Now assume that the only non-zero currents are $j_{a}^{0\nu}$ and $k_{a}^{0\nu}$;
$\Lambda=\Lambda(x_{5})$ and take the lagrangian as
$$L = [-(1/12) F^{\mu\nu}_{a} F_{a\enskip\mu\nu} - (1/12) G^{\mu\nu}_{a} G_{a\enskip\mu\nu}
- (1/4) j^{\mu}_{a} A_{a\enskip\mu} - (1/4) k^{\mu}_{a} B_{a\enskip\mu}$$
$$+ i\enskip f(\Lambda)\enskip \alpha\enskip \rho^{bc}\enskip 
A^{\mu}_{a}\enskip B_{a\enskip\mu}\enskip j^{\nu}_{b}\enskip k_{c\enskip\nu}]\enskip \Lambda(x)\eqno(11)$$
where the matrix $\rho_{ab}=\left(\matrix{0&1\cr 1&0\cr}\right)$. It is then
straigtforward to obtain  the dyon quantisation condition by proceeding exactly
as before and the result is 
$$e_{1} g_{2}+ e_{2}g_{1} = 2 \pi n\hbar \eqno(12)$$
Thus the quantisation condition depends on whether $p$ is odd or even.

In fact,the above procedure can be generalised to arbitrary $p$-form fields
by constructing appropriate field strengths $F,G$ and choosing the lagrangian as
$$L = [-(1/2)(1/(p+1)!) F^{\mu_{1}..\mu_{p+1}}_{a} F_{a\enskip\mu_{1}..\mu_{p+1}}- (1/2)(1/(p+1)!) G^{\mu\nu}_{a} G_{a\enskip\mu\nu}$$
$$- (1/2)(1/p!) j^{\mu_{1}..\mu_{p}}_{a} A_{a\enskip \mu_{1}..\mu_{p}} - (1/2)(1/p!) k^{\mu_{1}..\mu_{p}}_{a} B_{a\enskip \mu_{1}..\mu_{p}}$$
$$+ i\enskip f(\Lambda)\enskip \alpha\enskip \Omega^{bc}\enskip 
A^{\mu_{1}..\mu_{p}}_{a}\enskip B_{a\enskip \mu_{1}..\mu_{p}}\enskip j^{\mu_{1}..\mu_{p}}_{b}\enskip k_{c\enskip \mu_{1}..\mu_{p}}]\enskip \Lambda(x)\eqno(13)$$
where the matrix 
$$\Omega_{ab}=(1/2)[(1+(-1)^{p+1})\epsilon_{ab}\enskip +\enskip (1+(-1)^{p})\rho_{ab}]\eqno(14)$$
Currents will be defined as  
$$j^{\mu_{1}..\mu_{p}}_{a}=e_{a}\int dx^{\mu_{1}}\wedge dx^{\mu_{2}}....\wedge dx^{\mu_{p}}\eqno(15a)$$
$$k^{\mu_{1}..\mu_{p}}_{a}=g_{a}\int dx^{\mu_{1}}\wedge dx^{\mu_{2}}....\wedge dx^{\mu_{p}}\eqno(15b)$$
Assuming as before that the only non vanishing currents are
$j^{0\mu_{1}..\mu_{p-1}}_{a}$ and $k^{0\mu_{1}..\mu_{p-1}}_{a}$
and $\Lambda = \Lambda(x_{i})$ ,where $i$ is some spatial coordinate, one can solve
the relevant equations to get the dyon quantisation condition again. Depending
on whether $p$ is odd or even we will have
$$e_{1} g_{2} + (-1)^{p} e_{2} g_{1} = 2 \pi n\hbar \eqno(16)$$
We mention that for odd $p$ we will have anti-selfdual field strengths, while for
even $p$ we will have selfdual field strengths.

In conclusion, the spacetime dependent lagrangian formalism in conjunction
with the 't Hooft-Polyakov results have yielded an expression for the 
vacuum expectation value of the Higgs field as $A/e$. This result is definitely
susceptible to experiments. We have also shown that the 't Hooft ansatz for the
gauge field is sufficient to obtain an expression for the Higgs field if one
uses our formalism. No additional ansatz for $\phi$ is necessary.
The expression obtained reduces to the 't Hooft ansatz for
the Higgs field at $r\rightarrow\infty$. Finally, we have shown that classical solutions
of Yang-Mills theory also contain the germ of non-commuting coordinates 
residing on the boundary. The structure of these coordinates are like 
gauge fields and hence are relevant in constructing $D$-brane actions.

In conclusion, we have shown that the spacetime dependent lagrangian formulation
of electromagnetic duality can also accommodate the results of [3]. The dependence
of the quantisation condition on $p$ [3,4] is also accommodated. In our scheme
this has to do with the fact that coupling in the  interaction lagrangian depends
on $p$ through the matrix $\Omega_{ab}$. The importance of the dyon charge
quantisation in the theory of $D$-branes have been exhaustibly studied in [3].
So we do not elaborate on this. However, our formalism provides an alternate
interaction lagrangian picture of the same. The holographic principle [5] is again
illustrated-----the finite behaviour of $\Lambda$ on the boundary gives rise to
the exotic solutions within the bulk volume.

$^{*}$Electronic address : rajsekhar@vsnl.net

$^{\dagger}$Electronic address : debashis@boson.bose.res.in


\begin{thebibliography}{8}
\bibitem [1] {kn:xx}R.Bhattacharyya and D.Gangopadhyay, Mod.Phys.Lett. {\bf A15}, 901 (2000).
\bibitem [2] {kn:xx}R.Bhattacharyya and D.Gangopadhyay, Mod.Phys.Lett. {\bf A17}, 729 (2002);
D.Gangopadhyay,R.Bhattacharyya,L.P.Singh,{\it Spacetime Dependent Lagrangians
and the Barriola-Vilenkin Monopole Mass, hep-th/0208097}.
\bibitem [3] {kn:xx}P.Goddard and D.Olive, Rep.Prog.Phys. {\bf 41}, 1357 (1978).
\bibitem [4] {kn:xx}S.Coleman, {\it Phys.Rev.} {\bf D11} (1975) 2088;
S.Mandelstam, {\it Phys.Rev.} {\bf D11} (1975) 3026.
\bibitem [5] {kn:xx}N.Dadhich, Mod.Phys.Lett. {\bf A14}, 337 (1999);
\bibitem [6] {kn:xx}M.Barriola and A.Vilenkin, Phys.Rev.Lett. {\bf 63}, 341 (1989).
\bibitem [7] {kn:xx}E.Corrigan and D.Olive, Nucl.Phys. {\bf B110},237 (1976);
E.Corrigan, D.Olive and J.Nuyts, Nucl.Phys. {\bf B106}, 475 (1976).
\bibitem [8] {kn:xx}J.Polchinski,{\it String Theory, Vol 2},Cambridge Univ.Press,1998.
\bibitem [9] {kn:xx} G.t'Hooft,{\it Dimensional Reduction in Quantum Gravity,gr-qc/9310006};
L.Susskind, {\it Phys.Rev.} {\bf D49} (1994) 6606;
J.D.Bekenstein,{\it Phys.Rev.} {\bf D49} (1994) 1912;
J.Maldacena, {\it Adv.Theor.Math.Phys.} {\bf 2} (1998) 231;
E.Witten, {\it Adv.Theor.Math.Phys.} {\bf 2} (1998) 253; 
L.Susskind and E.Witten,{\it The Holographic bound in Anti-de Sitter Space, hep-th/9805114};
O.Aharony, S.S.Gubser,J.Maldacena,H.Ooguri and Y.Oz, {\it Large N Field Theories,
String Theory and Gravity, hep-th/9905111}; 
N.Seiberg and E.Witten, {\it String theory and noncommutative geometry,hep-th/9908142}.
\bibitem [10] {kn:xx}S.Deser, A.Gomberoff,M.Henneaux, Nucl.Phys. {\bf B520}, 179 (1998).
\bibitem [11] {kn:xx}S.Deser, M.Henneaux, A.Schwimmer, Phys. Lett. {\bf B428}, 284 (1998).

\end{thebibliography}
\end{document}